\documentclass[letterpaper,twocolumn,english]{revtex4}
\usepackage[T1]{fontenc}
\usepackage[latin9]{inputenc}
\usepackage{color}
\usepackage{graphicx}
\usepackage{amssymb}

\usepackage{babel}

\begin{document}

\title{Accumulation of Microswimmers due to Their Collisions with a Surface}

\author{Guanglai Li and Jay X. Tang}

\affiliation{Physics Department, Brown University, Providence RI 02912}
\begin{abstract}
In this letter we propose a kinematic model to show how collisions
with a surface and rotational Brownian motion give rise to the accumulation
of micro-swimmers near a surface. In this model, an elongated microswimmer
invariably travels parallel to the surface after hitting it from any
incident angle. It then swims away from the surface after some time,
facilitated by rotational Brownian motion. Simulations based on this
model reproduce the density distributions measured for the small bacteria
\textit{E. coli} and \textit{Caulobacter crescentus}, as well as for
the much larger bull spermatozoa swimming in confinement.
\end{abstract}


\maketitle
\textcolor{black}{Swimming aids the function and development of microorganisms
in many ways. For example, it enhances the formation of biofilms,
which provide favorable microenvironments for bacteria to cope with
environmental stresses~\cite{Watnick2000-2675}. Swimming also helps
transport spermatozoa toward eggs for fertilization~\cite{Levitan1995-228}.
Interestingly, the number density of cells as a function of distance
from a surface has been measured for }\textit{\textcolor{black}{E.
coli}}\textcolor{black}{~\cite{Berke2008-} and bull spermatozoa~\cite{Rothschild1963-1221},
showing in both cases values much higher near the surface than far
away. This near surface accumulation has mainly been attributed to
a hydrodynamic attraction between the cells and the surface~\cite{Fauci1995-679,Rothschild1963-1221}.
Recently, Berke }\textit{\textcolor{black}{et al.}}\textcolor{black}{~\cite{Berke2008-}
combined the effects of the hydrodynamic attraction and the translational
Brownian motion of the cells to predict the distribution of }\textit{\textcolor{black}{E.
coli}}\textcolor{black}{~as a function of distance. As noted by the
authors~\cite{Berke2008-}, however, this interpretation is not applicable
to cells within 10 $\mu$m from the surface, where most accumulation
occurs. }

\textcolor{black}{In this letter we present a different account for
the near surface accumulation. We ignore the hydrodynamic attraction
but emphasize the role of the collision with a surface at low Reynolds
number and rotational Brownian motion in a confined environment. We
show that a typical microswimmer with an elongated shape will swim
parallel to a surface after hitting it and therefore accumulate near
the surface. Rotational Brownian motion~\cite{Berg1993-Random} then
relaxes the accumulation by randomly changing the swimming direction
so that the cells have chances to swim away from the surface. In the
extreme case of no rotational Brownian motion, all the cells would
end up swimming in close proximity with the surface. In the opposite
extreme of very fast rotational Brownian motion, the cells will quickly
change to any possible swimming direction and subsequently would be
found anywhere with equal probability. In reality, a microorganism
randomly changes its swimming direction with a finite rotational diffusion
constant, resulting in a distribution in between the two extremes,
that is, more cells stay near the surface and fewer far away.}

We used the bacterium \textit{C. crescentus} strain CB15 SB3860, which
is a CheR mutant of $\Delta pilin$ (YB375) and swims forward exclusively,
to examine the details of near surface swimming. Swarmer cells of
this mutant do not follow circular trajectories when swimming near
surfaces~\cite{Li2008-18355}. The strains were synchronized with
the plate releasing method~\cite{Degnen1972-671,Li2008-18355} to
obtain cultures with primarily swimming cells. \textcolor{black}{The
synchronized culture was sealed between a glass slide and a coverslip
with vacuum grease for optical microscopy observation. Broken coverslip
pieces were used as spacers so that the thickness of the microscopy
sample is $\sim$200 $\mu$m. A 20$\times$ objective (Nikon Plan
Apo, NA 0.75) was used on a Nikon E800 microscope to take 5 snap shots
of swimming cells at 0.1 second intervals under darkfield mode using
a CoolSnap CCD camera (Princeton Instruments) and MetaMorph software
(Universal Imaging). The objective was focused on planes 5, 15, 25
$\mu$m, }\textit{\textcolor{black}{etc.}}\textcolor{black}{, away
from the surface and the cell number distribution was measured following
the method of Berke }\textit{\textcolor{black}{et al.}}\textcolor{black}{~\cite{Berke2008-}.
We noted that although this objective has a 1.4 $\mu$m depth of field,
cells up to nearly 5 $\mu$m off the focal plane appeared as bright
spots. Therefore the measured cell density was an average over a $\sim$10
$\mu$m thick layer.}

The swimming speed and the rotational diffusion constant were obtained
from the videos taken for cells over 50 $\mu$m away from both surfaces.
The average swimming speed was $\sim$45 $\mu$m/s. The rotational
diffusion constant was measured from $\sim$200 swimming trajectories.
The swimming direction at moment $t$ was taken as the direction from
the position at $t$ to the position at $t+0.1$ s. With this definition
the change in direction $\Delta\varphi$ over time interval $\Delta t$
was obtained and the rotational diffusion constant $D_{r}$ was calculated
to be 0.12 rad$^{2}$/s, using the equation $<\Delta\varphi^{2}>=2D_{r}\Delta t$.

\begin{figure}
\centering{}\includegraphics[width=7cm]{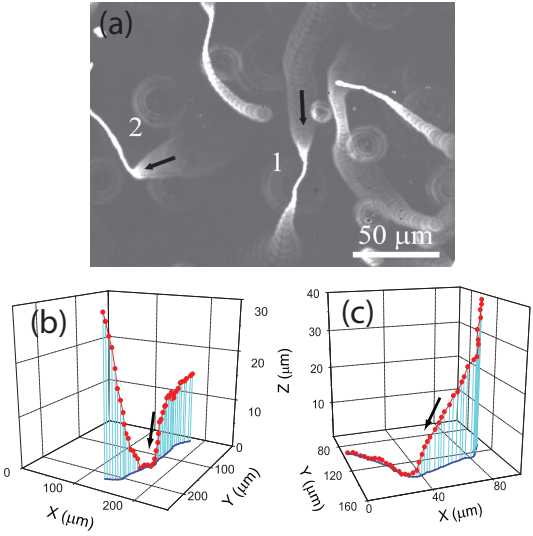}\caption{\label{fig:measurement}Trajectories of \textit{Caulobacter} swarmer
cells swimming near a glass surface. (a) Overlay of consecutive darkfield
images taken at 10 frames per second. (b) and (c) are 3-D plots (red)
and projections (blue) on the glass surface of the trajectories of
cells 1 and 2 in (a). Arrows indicate the swimming directions.}

\end{figure}

With particular interest we examined 3-D trajectories as the cells
approached and swam near a surface, until they took off. To do so,
we focused the objective on the top surface and recorded the swimming
trajectories. Example trajectories are shown in Fig.~\ref{fig:measurement}a
by overlaying consecutive frames taken at the rate of 10 frames per
second. The cell body appeared as a sharp bright spot when it was
in the focal plane and as a ring when it was away. Wu \textit{\textcolor{black}{et
al.}}~\cite{Wu2005-461} found that the ring size was proportional
to the distance of the cell from the focal plane and therefore can
be calibrated to determine the distance. Two examples of 3-D trajectories
of the cells 1 and 2 in Fig.~\ref{fig:measurement}a are plotted
in Figs.~\ref{fig:measurement}b and 1c. Most cells approached the
surface at an angle and then swam parallel to the surface for some
time before leaving. The manner of \textit{C. crescentus} hitting
a surface is similar to that of \textit{E. coli} observed with three
dimensional tracking microscopy~\cite{Frymier1995-6195}.

\begin{figure}
\begin{centering}
\includegraphics[width=6cm]{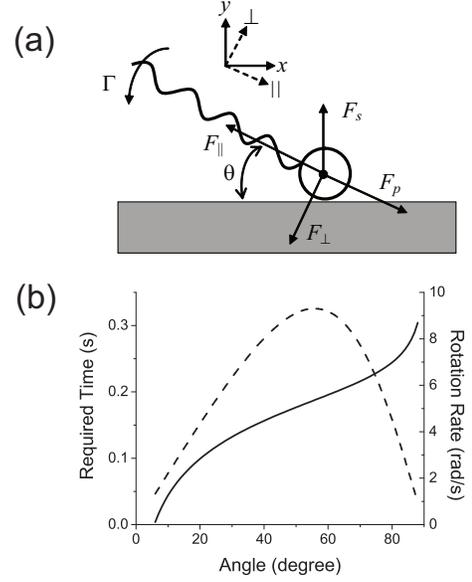}
\par\end{centering}

\caption{\label{fig:Forces}(a) Force and torque analysis of a forward swimming
cell hitting a surface. (b) Required time (solid) for the cell to
become parallel to the surface and the rotation rate (dashed) as function
of angle $\theta$.}

\end{figure}

We analyzed the force and torque on \textit{C. crescentus} swimming
near a surface and found that it would invariabley swim parallel to
the surface shortly after hitting the surface. For purposes of the
model, we approximate the cell as a sphere attached with a helical
filament of length $L$ propelled by a longitudinal force $F_{p}$.
After the cell hits the surface at an angle $\theta$, its velocity
component along the direction normal to the surface ($y$-axis Fig.~\ref{fig:Forces}a)
becomes zero. It will have a swimming speed $V_{x}$ along the $x$-axis
and a rotation rate $\Omega$ along the $z$-axis (not shown in the
figure). We ignore the increase in hydrodynamic drag on the cell due
to the nearby surface~\cite{Lauga2006-400,Li2008-18355} and assume
that the surface only provides a force $F_{s}$ to stop the swimming
along the $y$-axis. The hydrodynamic drag forces on the whole cell
(sphere plus helical filament) are split into components parallel
and perpendicular to the long axis, $F_{\parallel}$ and $F_{\perp}$.
The hydrodynamic torque $\Gamma$ on the whole cell is depicted with
respect to the sphere center. The forces and torque are given by \begin{equation}
\left(\begin{array}{c}
F_{\parallel}\\
F_{\perp}\\
\Gamma\end{array}\right)=\left(\begin{array}{ccc}
-A_{11} & 0 & 0\\
0 & -A_{22} & A_{23}\\
0 & A_{32} & -A_{33}\end{array}\right)\left(\begin{array}{c}
V_{\parallel}\\
V_{\perp}\\
\Omega\end{array}\right)\label{eq:force-torque}\end{equation}
where $V_{\parallel}=V_{x}\cos\theta$ and $V_{\perp}=V_{x}\sin\theta$
are the speed components along and perpendicular to the helical axis
and $A$ is the friction matrix, where $A_{ij}>0$ and $A_{23}=A_{32}$.

At a low Reynolds number, the force balance along $x$-axis is $F_{p}\cos\theta+F_{\parallel}\cos\theta+F_{\perp}\sin\theta=0$
and torque balance along $z$-direction is $\Gamma=0$, which together
determine the swimming speed and rotation rate as \begin{equation}
V_{x}=\frac{A_{33}\cos\theta}{A_{33}(A_{11}\cos^{2}\theta+A_{22}\sin^{2}\theta)-A_{23}^{2}\sin^{2}\theta}F_{p}\label{eq:v}\end{equation}
\begin{equation}
\Omega=\frac{A_{23}\sin\theta\cos\theta}{A_{33}(A_{11}\cos^{2}\theta+A_{22}\sin^{2}\theta)-A_{23}^{2}\sin^{2}\theta}F_{p}\label{eq:w}\end{equation}
Since $A_{22}A_{33}>A_{23}^{2}$, the denominator in the expressions
above are always positive. In the case as shown in Fig.~\ref{fig:Forces}a,
$V_{x}>0$ and $\Omega>0$. Therefore the cell swims toward the right
and the filament rotates toward the surface. 

We can estimate how fast the cell turns parallel to the surface during
the collision. Mathematically, the cell would take an infinitely long
time to become parallel to the surface, as calculated from Eq.~\ref{eq:w}.
In practice, however, since the rotational Brownian motion of \textit{C.
crescentus} varies its orientation by 0.1 rad within less than 0.1
sec, we estimate instead the time needed for the cell alignment with
the surface to fall under 0.1 rad. The parameters for a typical \textit{C.
crescentus~}\cite{Koyasu1984-125,Li2008-18355} are $A_{11}=2.2\times10^{-8}$
Nsm$^{-1}$, $A_{22}=2.5\times10^{-8}$ Nsm$^{-1}$, $A_{33}=1.9\times10^{-19}$
Nms, and $A_{23}=5.3\times10^{-14}$ Ns. The propulsive force is $F_{p}=A_{11}V\sim1\times10^{-12}$
N, where $V$ is the bulk swimming speed. The rotation rate after
hitting a surface is shown in Fig.~\ref{fig:Forces}b, which reaches
9 rad/s at 55$^{\circ}$. If a cell hits the surface at an angle $\theta_{0}$,
the time for it to become parallel is $\int_{0.1}^{\theta_{0}}d\theta/\Omega$
(Fig.~\ref{fig:Forces}b). This is less than 0.2 s for a typical
angle of $\theta_{0}=30^{\circ}$, and less than 0.3 s for an angle
as large as 85$^{\circ}$. Therefore in the following discussion we
state in a practical sense that a cell becomes parallel to the surface
after the collision. 

\begin{figure}
\begin{centering}
\includegraphics[width=6cm]{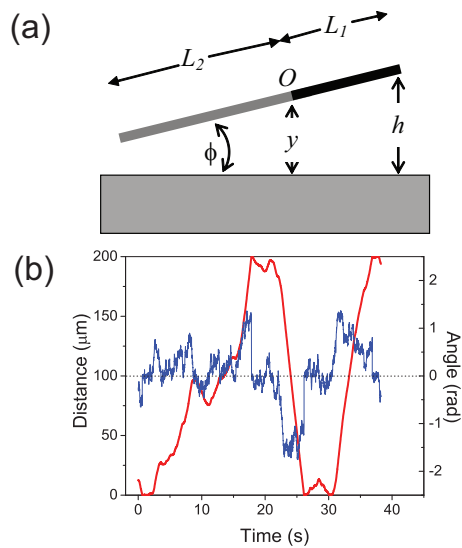}
\par\end{centering}

\caption{\label{fig:rod}(a) Rod model of a microswimmer near surface. The
black end represents that of the cell body and the gray end the flagellar
filament. (b) An example of simulated distance $h$ (red) and angle
$\phi$ (blue) as functions of time for the microswimmer, using the
parameters of \textit{C. crescentus}. The two surfaces are separated
by 200 $\mu$m.}

\end{figure}

Now we examine how a swimming microorganism takes off after hitting
a surface. To further simplify the model, we approximate the elongated
swimmer propelled by a longitudinal force as a nonuniform rod (Fig.~\ref{fig:rod}a).
This rod swims forward at speed $V$ in the bulk fluid. The rod has
a rotation center at position $O$, which is of a distance $L_{1}$
away from the head and $L_{2}$ away from the tail. The head has a
larger drag per unit length than the tail does, and thus $L_{1}<L_{2}$.
Due to the small size, the rod undergoes constant Brownian motion
with a rotational diffusion constant $D_{r}$ and translational diffusion
constant $D_{t}$. Since $A_{11}\sim A_{22}$, we ignore the angle
dependence of $D_{t}$. 

The change in distance of the rotation center to the surface $y$
is determined by the translational Brownian motion and the swimming
direction, which is constantly altered by the rotational Brownian
motion. Over a time interval $\Delta t$, $\Delta y=V\sin\phi\Delta t+\zeta\sqrt{2D_{t}\Delta t}$,
and $\Delta\phi=\varsigma\sqrt{2D_{r}\Delta t}$, where $\zeta$ and
$\varsigma$ are random numbers with zero mean and unit variance.
The translational Brownian motion contributes much less than swimming
to the displacement for microorganisms swimming at tens of $\mu$m/s.
When near the surface, the changes in distance and angle are also
restricted by the solid surface to satisfy $y\geqslant L_{1}\sin(-\phi)$
when the head is closer to the surface and $y\geqslant L_{2}\sin\phi$
when the tail is closer. Similar restrictions hold when a cell is
near the top surface. Knowing $D_{t}$ and $D_{r}$, we can track
the distance $y$ and angle $\phi$ over time. The distance of the
head from the surface $h$, which is what was measured in the experiments,
is determined by $h=y+L_{1}\sin\phi$. The probability distribution
of a cell at distance $h$ is obtained by tracking a cell swimming
between the two surfaces over 10$^{6}$ - 10$^{7}$ sec. 

\begin{figure}
\begin{centering}
\includegraphics[width=8cm]{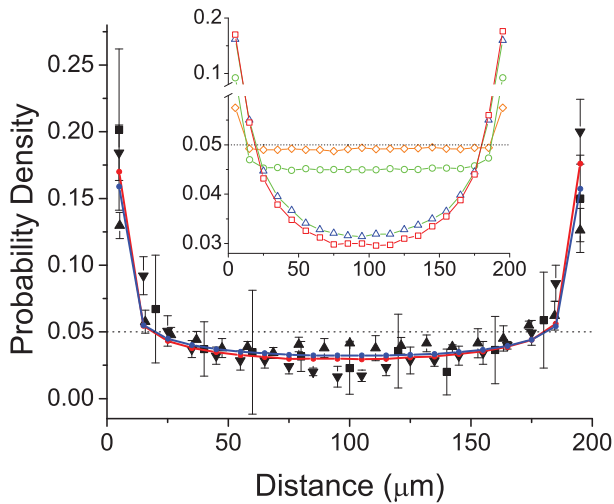}
\par\end{centering}

\caption{\label{fig:Different Dr}Comparison between simulated density distributions
at rotational diffusion constants 0.12 (blue) and 0.0001 (red) rad$^{2}$/s
and the measured distributions of \textit{C. crescentus} (up triangle),
\textit{E. coli} (down triangle, ref~\cite{Berke2008-}), and bull
spermatozoa (square, ref~\cite{Rothschild1963-1221}). The separation
between the two glass surfaces is 200 $\mu$m. Inset compares simulated
distribution at rotational diffusion constants of 10 (diamond), 1
(circle), 0.1 (triangle), and 0.0001 (square) rad$^{2}$/s at a swimming
speed of 50 $\mu$m/s, corresponding to rod lengths of $\sim$1.3,$\sim$2.8,
$\sim$6, and $\sim$60 $\mu$m, respectively. The dotted lines indicate
the probability density if there is no surface accumulation.}

\end{figure}

We simulated the distance and angle of swimming \textit{C. crescentus}
between two glass surfaces separated by 200 $\mu$m. The cell was
treated as a $L=6$ $\mu$m rod, with a typical translational diffusion
constant on the order of 0.1 $\mu$m$^{2}$/s and the measured rotational
diffusion constant of 0.12 rad$^{2}$/s. The rotation center was approximated
at a position where $L_{1}=0.3L$. Fig.~\ref{fig:rod}b shows examples
of distance (red) and angle (blue) varying over time. The cell hits
the top and bottom surfaces repeatedly as it swims between them.\textcolor{black}{~The
simulated distance from the bottom surface was recorded every 0.1
second and a histogram of distances was made using a bin size of 10
$\mu$m.~The simulated distribution is plotted in Fig.~\ref{fig:Different Dr}
(blue) and compared with the measured one for }\textit{\textcolor{black}{C.
crescentus}}\textcolor{black}{~(triangle). The simulation clearly
shows higher densities near the surfaces, with the entire profile
in excellent agreement with the measurements.}

This model is also applicable to the distribution of \textit{E. coli}
and bull spermatozoa between two surfaces. We took the cell number
distribution of \textit{E. coli} from reference~\cite{Berke2008-}
and that of bull spermatozoa from reference~\cite{Rothschild1963-1221},
converted them to probability density, and plotted them in comparison
with that of \textit{C. crescentus} in Fig.~\ref{fig:Different Dr}.
\textit{E. coli} is similar in size to \textit{C. crescentus} and
it is not surprising that they have similar distributions. Bull spermatozoa
are ten times larger yet surprisingly the distribution is similar
to that of bacteria. Nevertheless, this similarity is actually predicted
by our model. To simulate for bull spermatozoa, we treated it as a
60 $\mu$m long rod with a translational diffusion constant on the
order of 0.01 $\mu$m$^{2}$/s and a rotational diffusion constant
of $D_{r}=10^{-4}$ rad$^{2}$/s, which is $\sim$1000 times smaller
than that of \textit{C. crescentus}. The simulation shows that the
difference in distribution between the bull spermatozoa (red) and
the \textit{C. crescentus} (red) is so small that it cannot be distinguished
by the observations under the set conditions.

\textcolor{black}{The distribution of swimming cells near a surface
can be understood intuitively based on the variation of angle due
to rotational Brownian motion. Once a cell hits a surface, the angle
becomes zero and the cell swims parallel to the surface for some time.
The cell then swims away from the surface by changing swimming direction
due to rotational Brownian motion. Let us examine, for instance, the
time $t_{1}$ and $t_{2}$ needed for a cell to swim to different
distances from the surface, $h_{1}<h_{2}$. Since the cell swims to
distance $h_{1}$ before reaching $h_{2}$, $t_{1}<t_{2}$ must hold.
Assuming the swimming direction is only affected by rotational Brownian
motion, the mean square angle at time $t$ is $<\phi^{2}>\sim2D_{r}t$.
Therefore, the closer a cell swims near the surface, the smaller the
angle is. The perpendicular component of swimming speed $V_{y}=V\sin\phi$
is smaller when closer to the surface. The dwell time of a cell staying
at a distance $h$ is inversely proportional to $V_{y}$ and therefore
the probability density is larger near the surface. Additional features
of the distribution can be understood based on this physical picture.
When the cell swims close to the surface after a collision, $\phi\ll1$.
The probability density is $\sim1/\phi$, which drops sharply with
the increasing angle and distance. This picture also predicts that
a microswimmer with a large rotational diffusion constant leaves a
surface more rapidly after a collision and hence the accumulation
near the surface is weaker, as shown for $D_{r}=10$ rad$^{2}$/s
in the inset of Fig.~\ref{fig:Different Dr}. }

\textcolor{black}{In this model we have ignored the hydrodynamic interaction
between the cell and the surface. In reality, the swimming cell generates
a flow, which interacts with the nearby surface, reorienting and attracting
the cell towards the surface. Berke }\textit{\textcolor{black}{et
al.}}\textcolor{black}{~\cite{Berke2008-} calculated this effect
for bacteria when the cell is $>$10 $\mu$m away from the surface.
A simple estimation shows that this effect is small compared to rotational
Brownian motion and swimming when the cell is nearly parallel to the
surface. For example, at a distance $h=10$ $\mu$m and angle $\phi=0.1$
rad, calculation based on their model yields a reorientation rate
of $\sim$0.01 rad/s and an attraction speed of $\sim$1 $\mu$m/s,
while in 1 second the rotational Brownian motion can reorient the
cell by 0.4 rad on average and the component of swimming speed normal
to the surface is on the order of 10 $\mu$m/s. Therefore, the modification
to the distribution due to the long range hydrodynamic interaction
is small when the distance is $>$10 $\mu$m. When a cell is less
than 1 $\mu$m from the surface, the large hydrodynamic friction between
the cell and surface may keep the cells near the surface for a long
time~\cite{Lauga2006-400}. The effect of this extension of dwell
time is not that dramatic since the distribution is binned by 10 $\mu$m
in distance. The hydrodynamic interaction in the range of 1 to 10
$\mu$m is yet to be described theoretically. Its effect on distribution
of cells in this range, however, is expected to be secondary as evident
by the good agreement between the simulation results ignoring it and
the experimental measurements.}

In conclusion, we have demonstrated the effects of collision and rotational
Brownian motion on swimming microorganisms near surfaces. The collision
with a surface resets the swimming direction to be parallel to it
and the rotational Brownian motion then randomly alters the swimming
direction, which leads to the accumulation. An excellent agreement
is obtained between the simulations based on this picture and the
experimental results. Recently, various artificial microswimmers have
been developed~\cite{Dreyfus2005-862,Ogrin2008-,Dhar2006-66}, which
usually have an elongated shape. Similar effect of the collision and
rotational Brownian motion is expected when they swim near a surface. 

This work is supported by NIH GM077648 and NSF CMMI 0825873. We thank
Professors Y. Brun of Indiana University and B. Ely at University
of South Carolina for providing the bacteria strains.

\bibliographystyle{apsrev}
\addcontentsline{toc}{section}{\refname}
%\bibliography{bibtext_master}

\end{document}